\documentclass[%
 amsmath,amssymb,
 aps,
prd,
twocolumn, superscriptaddress
]{revtex4-1}

\usepackage[normalem]{ulem}

\usepackage{graphicx}
\usepackage{dcolumn}
\usepackage{bm}
\usepackage{hyperref}

\usepackage{xcolor}

\begin{document}

\preprint{APS/123-QED}

\title{Moment of inertia of slowly rotating anisotropic neutron stars in $f(R,T)$ gravity}

\author{Juan M. Z. Pretel}
 \email{juanzarate@cbpf.br}
 \affiliation{
 Centro Brasileiro de Pesquisas F{\'i}sicas, Rua Dr.~Xavier Sigaud, 150 URCA, Rio de Janeiro CEP 22290-180, RJ, Brazil
}

\date{\today}

\begin{abstract}
Within the framework of $f(R,T)$ theories of gravity, we investigate the hydrostatic equilibrium of anisotropic neutron stars with a physically relevant equation of state (EoS) for the radial pressure. In particular, we focus on the $f(R,T) = R+ 2\beta T$ model, where $\beta$ is a minimal coupling constant. In the slowly rotating approximation, we derive the modified TOV equations and the expression for the relativistic moment of inertia. The main properties of neutron stars, such as radius, mass and moment of inertia, are studied in detail. Our results revel that the main consequence of the $2\beta T$ term is a substantial increase in the surface radius for low enough central densities. Nevertheless, such a term slightly modifies the total gravitational mass and moment of inertia of the slowly rotating stars. Furthermore, the changes are noticeable when anisotropy is incorporated into the stellar fluid, and it is possible to obtain higher masses that are consistent with the current observational data.   
\end{abstract}

\maketitle


\section{Introduction} 

Despite the great success of General Relativity (GR) in predicting various gravitational phenomena tested in the solar system \cite{Will2014} and in strong-field situations (such as the final stage of compact-object binaries \cite{Abbott221101, Abbott011102}), it could not help to identify the nature of dark energy and other puzzles. In other words, there are still many open problems in modern cosmology and it is well known that GR is not the only theory of gravity \cite{Saridakis2021}. Indeed, it has been shown that GR is not renormalizable as a quantum field theory unless higher-order curvature invariants are included in its action \cite{Stelle1977, Vilkovisky1992}. Furthermore, GR requires modifications at small time and length scales or at energies comparable with the Planck energy scales. In that regard, it has been argued that the early-time inflation and the late-time accelerated expansion of the Universe can be an effect of the modification of the geometric theory formulated by Einstein \cite{Starobinsky1980, Capozziello2002, Carroll2004, Nojiri2007}.

One of the simplest ways to modify GR is by replacing the Ricci scalar $R$ in the standard Einstein-Hilbert action by an arbitrary function of $R$, this is, the so-called $f(R)$ theories of gravity \cite{SotiriouFaraoni, Felice}. Extensive and detailed reviews on the cosmological implications of such theories can be found in Refs.~\cite{Capozziello2011, Nojiri2011, Clifton2012, Nojiri2017}. On the other hand, at astrophysical level, these theories basically change the Tolman-Oppenheimer-Volkoff (TOV) equations and hence the astrophysical properties of compact stars, such as mass-radius relations, maximum masses, or moment of inertia are somehow altered. See Ref.~\cite{Olmo2020} for a broad overview about relativistic and non-relativistic stars within the context of modified theories of gravity formulated in both metric and metric-affine approaches.

In most of the works reported in the literature about internal structure of compact stars in GR and modified theories of gravity it is very common to assume that such stars are made up of an isotropic perfect fluid. Nevertheless, there are strong arguments indicating that the impact of anisotropy (this is, unequal radial and tangential pressures) cannot be neglected when we deal with nuclear matter at very high densities and pressures, for instance, see Refs.~\cite{HerreraSantos1997, Isayev2017, Ivanov2017, Maurya2018, Biswas2019, Pretel2020EPJC, Bordbar2022} and references therein. In that regard, it has been shown that the presence of anisotropy can lead to significant changes in the main characteristics of compact stars \cite{Maurya2018, Biswas2019, Pretel2020EPJC, Horvat2011, Rahmansyah2020, Roupas2020, Das2021, Das2021GRG, Roupas2021, Das2022EPJC}. Within the framework of extended theories of gravity, it is also important to mention that non-rotating anisotropic compact stars have been recently studied by some authors in Refs.~\cite{Shamir2017, Folomeev2018, Mustafa2020, Nashed2021A, Nashed2021B, Deb2019, Maurya2019, Biswas2020, Maurya2020, Rej2021, BISWAS2021168429, Vernieri2019, Mota2019, Ashraf2020, Tangphati2021, Tangphati2021PDU, NashedApJ, Solanki2022, Pretel2022}. In addition, in the context of scalar-tensor theory of gravity, slowly rotating anisotropic neutron stars have been investigated in Ref.~\cite{Silva2015}.

Harko and collaborators \cite{Harko2011} have proposed a generalization of $f(R)$ modified theories of gravity in order to introduce a coupling between geometry and matter, namely $f(R,T)$ gravity, where $T$ denotes the trace of the energy-momentum tensor. Indeed, the simplest and most studied model involving a minimal matter-gravity coupling is given by $f(R,T)= R+ 2\beta T$ gravity. The cosmological aspects of this model have been recently explored in Refs.~\cite{Shabani2018, Debnath2019, Bhattacharjee2020, Bhattacharjee2020EPJP, Gamonal2021}, while other authors have investigated the astrophysical consequences of the $2\beta T$ term on the equilibrium structure of isotropic \cite{Moraes2016, Das2016, Deb2018, Deb_2019JCAP, Lobato2020, Pretel2021, PretelCPC, Bora2022} and anisotropic \cite{Deb2019, Maurya2019, Biswas2020, Maurya2020, Rej2021, BISWAS2021168429} compact stars. A characteristic of this model is that $R=0$ outside a compact star, and hence the exterior spacetime is still described by the Schwarzschild exterior solution. As a result, it has been shown that for high enough central densities the contributions of the $2\beta T$ term are irrelevant, whereas below a certain central density value the radius of an isotropic compact star undergoes substantial deviations from GR \cite{Lobato2020, Pretel2021}.

To determine the equilibrium configurations and moment of inertia of slowly rotating anisotropic stars up to first order in the angular velocity, we will employ a physically motivated functional relation $\sigma$ (defined as the difference between radial and tangential pressure) for the anisotropy profile known in the literature as quasi-local ansatz \cite{Horvat2011}. Moreover, we will follow a procedure analogous to that carried out by Hartle in GR \cite{Hartle1967} in order to obtain the modified version of the differential equation which governs the difference between the angular velocity of the star and the angular velocity of the local inertial frames.

To achieve our results, the present work is organized as follows: In Sec. \ref{Sec2} we briefly review $f(R,T)$ gravity and we present the corresponding relativistic equations for the $f(R,T)= R+ 2\beta T$ model. In Sec. \ref{Sec3} we derive the modified TOV equations for anisotropic stellar configurations by adopting a non-rotating and slowly rotating metric. Section \ref{Sec4} presents a well-known EoS to describe neutron stars as well as the anisotropy ansatz. In Sec. \ref{Sec5} we discuss our numerical results, and finally, our conclusions are presented in Sec. \ref{Sec6}. In this paper we will use a geometric unit system and the sign convention $(-,+,+,+)$. However, our results will be given in physical units.



\section{Basic formalism of $f(R,T)$ gravity}\label{Sec2} 

A more general formulation of $f(R)$ modified theories of gravity consists in the inclusion of an explicit gravity-matter coupling by means of an arbitrary function of the Ricci scalar $R$ and the trace of the energy-momentum tensor $T$. Thus, the modified Einstein-Hilbert action in $f(R,T)$ gravity is given by \cite{Harko2011}
\begin{equation}\label{1}
    S = \frac{1}{16\pi}\int f(R,T)\sqrt{-g}d^4x + \int\mathcal{L}_m \sqrt{-g}d^4x ,
\end{equation}
where $g$ is the determinant of the spacetime metric $g_{\mu\nu}$ and $\mathcal{L}_m$ denotes the Lagrangian density for matter fields. The corresponding field equations in $f(R,T)$ gravity can be obtained from the variation of the action (\ref{1}) with respect to the metric:
\begin{align}\label{2}
    f_R(R,T) R_{\mu\nu} &- \dfrac{1}{2}f(R,T) g_{\mu\nu} + [g_{\mu\nu}\square - \nabla_\mu\nabla_\nu] f_R(R,T)  \nonumber  \\
    &= 8\pi T_{\mu\nu} -(T_{\mu\nu} + \Theta_{\mu\nu})f_T(R,T) ,
\end{align}
where $R_{\mu\nu}$ is the Ricci tensor, $T_{\mu\nu}$ the energy-momentum tensor, $f_R \equiv \partial f/\partial R$, $f_T \equiv \partial f/\partial T$, $\square \equiv \nabla_\mu\nabla^\mu$ is the d'Alembertian operator with $\nabla_\mu$ standing for the covariant derivative, and the tensor $\Theta_{\mu\nu}$ is defined in terms of the variation of $T_{\mu\nu}$ with respect to the metric, namely  
\begin{align}\label{3}
    \Theta_{\mu\nu} &\equiv g^{\alpha\beta}\frac{\delta T_{\alpha\beta}}{\delta g^{\mu\nu}}  \nonumber  \\
    &=  -2T_{\mu\nu} + g_{\mu\nu}\mathcal{L}_m - 2g^{\alpha\beta} \frac{\partial^2\mathcal{L}_m}{\partial g^{\mu\nu} \partial g^{\alpha\beta}} .
\end{align}

Just as in $f(R)$ gravity \cite{SotiriouFaraoni, Felice}, in $f(R,T)$ theories the Ricci scalar is also a dynamical entity which is described by a differential equation obtained by taking the trace of the field equations (\ref{2}), this is
\begin{align}\label{4}
    3\square f_R(R,T) &+ Rf_R(R,T) - 2f(R,T)  \nonumber  \\
    &= 8\pi T - (T+\Theta)f_T(R,T) ,
\end{align}
where we have denoted $\Theta= \Theta_\mu^{\ \mu}$. In addition, the four-divergence of Eq. (\ref{2}) yields \cite{Barrientos2014}
\begin{align}\label{5}
    \nabla^\mu T_{\mu\nu} =&\ \frac{f_T(R,T)}{8\pi - f_T(R,T)}\bigg[ (T_{\mu\nu} + \Theta_{\mu\nu})\nabla^\mu \ln f_T(R,T)   \nonumber  \\
    & + \nabla^\mu\Theta_{\mu\nu} - \frac{1}{2}g_{\mu\nu}\nabla^\mu T \bigg] .
\end{align}

In order to obtain numerical solutions that describe compact stars, one has to specify the particular model of $f(R,T)$ gravity. In that regard, we consider the simplest model involving a minimal matter-gravity coupling proposed by Harko {\it et al.}~\cite{Harko2011}, i.e. $f(R,T)= R+ 2\beta T$ gravity, which has been the most studied model of $f(R,T)$ gravity at both astrophysical and cosmological scale. As a consequence, Eqs.~(\ref{2}), (\ref{4}) and (\ref{5}) can be written as follows
\begin{align}
    G_{\mu\nu} &= 8\pi T_{\mu\nu} + \beta Tg_{\mu\nu} - 2\beta(T_{\mu\nu} + \Theta_{\mu\nu}) ,   \label{6}   \\ 
    R &= -8\pi T - 2\beta(T- \Theta) ,   \label{7}    \\
    \nabla^\mu T_{\mu\nu} &= \frac{2\beta}{8\pi - 2\beta} \left[ \nabla^\mu \Theta_{\mu\nu} - \frac{1}{2}g_{\mu\nu}\nabla^\mu T \right] ,   \label{8}
\end{align}
where $G_{\mu\nu}$ is the Einstein tensor.


\section{Modified TOV equations}\label{Sec3}

\subsection{Non-rotating stars}

We shall assume that the matter source is described by an anisotropic perfect fluid with energy density $\rho$, radial pressure $p_r$ and tangential pressure $p_t$. Under theses assumptions, the energy-momentum tensor is given by 
\begin{equation}\label{9}
T_{\mu\nu} = (\rho + p_t) u_\mu u_\nu + p_t g_{\mu\nu} - \sigma k_\mu k_\nu ,
\end{equation}
with $u^\mu$ being the four-velocity of the fluid and which satisfies the normalization property $u_\mu u^\mu = -1$, $k^\mu$ is a unit radial four-vector so that $k_\mu k^\mu = 1$, and $\sigma \equiv p_t - p_r$ is the anisotropy factor. 

In addition, we consider that the interior spacetime of the spherically symmetric stellar configuration is described by the standard line element
\begin{equation}\label{10}
ds^2 = -e^{2\psi}dt^2 + e^{2\lambda}dr^2 + r^2(d\theta^2 + \sin^2\theta d\phi^2) ,
\end{equation}
where $x^\mu = (t, r, \theta, \phi)$ are the Schwarzschild-like coordinates, and the metric potentials $\psi$ and $\lambda$ are functions only of the radial coordinate in a hydrostatic equilibrium situation. Consequently, we can write $u^\mu = e^{-\psi}\delta_0^\mu$, $k^\mu= e^{-\lambda}\delta_1^\mu$ and the trace of the energy-momentum tensor (\ref{9}) takes the form $T= -\rho + 3p_r+ 2\sigma$.

Within the context of anisotropic fluids in $f(R,T)$ gravity, the most adopted choice in the literature for the matter Lagrangian density is given by $\mathcal{L}_m = \mathcal{P}$, where $\mathcal{P} \equiv (p_r+ 2p_t)/3$. For more details about this choice, see Refs.~\cite{Deb2019, Maurya2019, Biswas2020, Maurya2020, BISWAS2021168429}. Under this consideration, $\Theta_{\mu\nu} = -2T_{\mu\nu} + \mathcal{P}g_{\mu\nu}$ and Eqs.~(\ref{6}), (\ref{7}) and (\ref{8}) become 
\begin{align}
    G_{\mu\nu} &= 8\pi T_{\mu\nu} + \beta Tg_{\mu\nu} + 2\beta(T_{\mu\nu} - \mathcal{P}g_{\mu\nu}) ,   \label{11}   \\ 
    R &= -8\pi T - 2\beta(3T- 4\mathcal{P}) ,   \label{12}    \\
    \nabla^\mu T_{\mu\nu} &= \frac{2\beta}{8\pi + 2\beta} \partial_\nu\left( \mathcal{P} - \frac{1}{2}T \right) .   \label{13}
\end{align}

For the metric (\ref{10}) and energy-momentum tensor (\ref{9}), the non-zero components of the field equations (\ref{11}) are explicitly given by
  \begin{align}
      & \frac{1}{r^2}\frac{d}{dr}(re^{-2\lambda}) - \frac{1}{r^2} = -8\pi\rho + \beta\left[ -3\rho + p_r + \frac{2}{3}\sigma \right] ,  \label{14}  \\
      & e^{-2\lambda}\left( \frac{2}{r}\psi' + \frac{1}{r^2} \right)- \frac{1}{r^2} = 8\pi p_r + \beta\left[ -\rho + 3p_r+ \frac{2}{3}\sigma \right],  \label{15}  \\
      & e^{-2\lambda}\left[ \psi''+ \psi'^2- \psi'\lambda'+ \frac{1}{r}(\psi'- \lambda') \right]   \nonumber  \\
      &\hspace{1.5cm} = 8\pi(p_r + \sigma) + \beta\left[ -\rho + 3p_r+ \frac{8}{3}\sigma \right] ,  \label{16}
  \end{align}
where the prime represents differentiation with respect to the radial coordinate. Moreover, Eq.~(\ref{13}) implies that
\begin{align}\label{17}
    \frac{dp_r}{dr} =& -(\rho + p_r)\psi'+ \frac{2}{r}\sigma  \nonumber \\
    &+ \frac{\beta}{8\pi +2\beta}\frac{d}{dr}\left[ \rho - p_r- \frac{2}{3}\sigma \right] .
\end{align}

Eq. (\ref{14}) leads to 
\begin{equation}\label{18}
    re^{-2\lambda} = r- \int r^2\left[ 8\pi\rho + \beta \left( 3\rho - p_r- \frac{2}{3}\sigma \right) \right]dr ,
\end{equation}
or alternatively, 
\begin{equation}\label{19}
    e^{-2\lambda} = 1 - \frac{2m}{r} ,
\end{equation}
where $m(r)$ represents the gravitational mass within a sphere of radius $r$, given by 
\begin{align}\label{20}
    m(r) =&\ 4\pi\int_0^r \bar{r}^2\rho(\bar{r})d\bar{r}  \nonumber  \\
    &+ \frac{\beta}{2}\int_0^r \bar{r}^2 \left[ 3\rho(\bar{r}) - p_r(\bar{r}) - \frac{2}{3}\sigma(\bar{r}) \right]d\bar{r} .
\end{align}

At the surface, where the radial pressure vanishes, $M \equiv m(r_{\rm sur})$ is the total mass of the anisotropic compact star. From our anisotropic version (\ref{20}), here we can see that by making $\sigma= 0$ one recovers the mass function for the isotropic case given in Ref.~\cite{Pretel2021}. In view of Eq.~(\ref{19}), from Eq.~(\ref{15}) we obtain 
\begin{align}\label{21}
    \psi'=& \left[ \frac{m}{r^2}+ 4\pi rp_r + \frac{\beta r}{2}\left( -\rho + 3p_r+ \frac{2}{3}\sigma \right) \right]  \nonumber  \\
    &\times \left( 1- \frac{2m}{r} \right)^{-1} ,
\end{align}
and hence the relativistic structure of an anisotropic compact star within the context of $f(R,T)= R+ 2\beta T$ gravity is described by the modified TOV equations:
\begin{align}
    \frac{dm}{dr} =&\ 4\pi r^2\rho + \frac{\beta r^2}{2}\left( 3\rho - p_r- \frac{2}{3}\sigma \right) ,  \label{22}  \\
    \frac{dp_r}{dr} =& -\frac{\rho + p_r}{1+a} \left[ \frac{m}{r^2}+ 4\pi rp_r + \frac{\beta r}{2}\left( 3p_r- \rho + \frac{2}{3}\sigma \right) \right]  \nonumber  \\
    &\times \left( 1- \frac{2m}{r} \right)^{-1} + \frac{a}{1+a}\frac{d\rho}{dr}  \nonumber  \\
    &+ \frac{2}{1+a}\left[ \frac{\sigma}{r} - \frac{a}{3}\frac{d\sigma}{dr} \right] ,  \label{23}  \\
    \frac{d\psi}{dr} =& \frac{1}{\rho+ p_r}\left[ -(1+a)\frac{dp_r}{dr} + a\frac{d\rho}{dr} + 2\left( \frac{\sigma}{r} - \frac{a}{3}\frac{d\sigma}{dr} \right)\right]  \label{24} , 
\end{align}
where we have defined $a \equiv \beta/(8\pi + 2\beta)$. As expected, the modified TOV equations in the isotropic scenario are retrieved when $p_r= p_t$ \cite{Pretel2021}. Furthermore, when the minimal coupling constant vanishes (this is, $\beta =0$), we can recover the standard TOV equations for anisotropic stars in GR \cite{Pretel2020EPJC}. 

Given an EoS for the radial pressure $p_r= p_r(\rho)$ and an anisotropy relation for $\sigma$, Eqs.~(\ref{22}) and (\ref{23}) can be integrated by guaranteeing regularity at the center of the star and for a given value of central energy density. In addition, according to Eq.~(\ref{12}), we notice that $R=0$ in the outer region of the star. This means that we can still use the Schwarzschild vacuum solution to describe the exterior spacetime so that the interior solution is matched at the boundary $r= r_{\rm sur}$ to the exterior Schwarzschild solution. Thus, the system of equations (\ref{22})-(\ref{24}) can be solved by imposing the following boundary conditions
\begin{align}\label{25}
m(0) &= 0,   &   \rho(0) &= \rho_c,   &   \psi(r_{\rm sur}) &= \frac{1}{2}\ln\left[ 1 - \frac{2M}{r_{\rm sur}} \right] . \ \
\end{align}

\subsection{Slowly rotating stars}

In the slowly rotating approximation \cite{Hartle1967}, i.e., when rotational corrections appear at first order in the angular velocity of the stars $\Omega$, the spacetime metric (\ref{10}) is replaced by its slowly rotating counterpart \cite{Hartle1967, Staykov2014} 
\begin{align}\label{26}
  ds^2 =& -e^{2\psi(r)}dt^2 + e^{2\lambda(r)}dr^2 + r^2(d\theta^2 + \sin^2\theta d\phi^2)  \nonumber  \\
  & -2\omega(r,\theta)r^2\sin^2\theta dtd\phi ,
\end{align}
where $\omega(r,\theta)$ stands for the angular velocity of the local inertial frames dragged by the stellar rotation. In other words, if a particle is dropped from rest at a great distance from the rotating star, the particle would experience an ever increasing drag in the direction of rotation of the star as it approaches. In fact, here it is convenient to define the difference $\varpi \equiv \Omega- \omega$ as the coordinate angular velocity of the fluid element at ($r,\theta$) seen by the freely falling observer \cite{Hartle1967}.

Since $\Omega$ is the angular velocity of the fluid as seen by an observer at rest at some spacetime point $(t,r,\theta,\phi)$, one finds that the four-velocity up to linear terms in $\Omega$ is given by $u^\mu= (e^{-\psi},0,0,\Omega e^{-\psi})$. To this order, the spherical symmetry is still preserved and it is possible to extend the validity of the TOV equations (\ref{22})-(\ref{24}). Nevertheless, the $03$-component of the field equations contributes an additional differential equation for angular velocity $\omega(r,\theta)$. By retaining only first-order terms in the angular velocity, we have $T_{03}= -[ \varpi(\rho+ p_t) + \omega p_t ]r^2\sin^2\theta$ and hence Eq.~(\ref{11}) gives the following expression
\begin{align}\label{27}
    G_{03} =& -\bigg[ 2(4\pi+ \beta)(\rho + p_t)\varpi + 8\pi\omega p_t   \nonumber  \\
    &\left. +\beta\left( -\rho+ \frac{1}{3}p_r + \frac{8}{3}p_t \right)\omega\right] r^2\sin^2\theta ,
\end{align}
or alternatively,
\begin{align}\label{28}
    &\frac{e^{\psi-\lambda}}{r^4}\frac{\partial}{\partial r}\left[ e^{-(\psi+\lambda)}r^4\frac{\partial\varpi}{\partial r} \right] + \frac{1}{r^2\sin^3\theta}\frac{\partial}{\partial\theta}\left[ \sin^3\theta\frac{\partial\varpi}{\partial\theta} \right]  \nonumber  \\
    &\hspace{2cm}= 4(4\pi+ \beta)(\rho+ p_t)\varpi . 
\end{align}

Following the procedure carried out by Hartle in GR \cite{Hartle1967} and Staykov \textit{et al.}~in $R^2$-gravity \cite{Staykov2014}, we expand $\varpi$ in the form 
\begin{equation}\label{29}
    \varpi(r,\theta) = \sum_{l=1}^\infty \varpi_l(r)\left( \frac{-1}{\sin\theta}\frac{dP_l}{d\theta} \right) ,
\end{equation}
where $P_l$ are Legendre polynomials. In view of Eq.~(\ref{29}), we can write
\begin{align}
 \frac{\partial}{\partial\theta}\left[ \sin^3\theta\frac{\partial\varpi}{\partial\theta} \right] =& \sum_l \varpi_l(r) \left[ (\cos^2\theta - \sin^2\theta)\frac{dP_l}{d\theta}  \right.  \nonumber  \\
 &\hspace{1cm}\left. - \sin\theta\cos\theta\frac{d^2P_l}{d\theta^2} - \sin^2\theta\frac{d^3P_l}{d\theta^3} \right]  \nonumber  \\
 =& \sum_l \varpi_l(r) \left[ l(l+1) -2 \right]\sin^2\theta\frac{dP_l}{d\theta} , \label{30}
\end{align}
where we have used the Legendre differential equation
\begin{equation}
  \frac{d^2P_l}{d\theta^2} + \frac{\cos\theta}{\sin\theta}\frac{dP_l}{d\theta} + l(l+1)P_l = 0 .
\end{equation}

Thus, after substituting Eqs.~(\ref{29}) and (\ref{30}) into (\ref{28}), we get
\begin{align}\label{32}
    &\frac{e^{\psi-\lambda}}{r^4}\frac{d}{dr}\left[ e^{-(\psi+\lambda)}r^4\frac{d\varpi_l}{dr} \right] - \frac{l(l+1)-2}{r^2}\varpi_l  \nonumber  \\
    &\hspace{2cm}= 4(4\pi+ \beta)(\rho+ p_t)\varpi_l . 
\end{align}

At great distances from the stellar surface, where spacetime must be asymptotically flat, the solution of Eq.~(\ref{32}) assumes the form $\varpi_l(r) \rightarrow c_1 r^{-l-2} + c_2r^{l-1}$. Furthermore, the dragging angular velocity is expected to be $\omega \rightarrow 2J/r^3$ (or alternatively, $\varpi \rightarrow \Omega - 2J/r^3$) for $r \rightarrow \infty$, where $J$ is the angular momentum carried out by the star (see Ref.~\cite{Glendenning} for more details). Therefore, by comparison we can see that all coefficients in the Legendre expansion vanish except for $l=1$. This means that $\varpi$ is a function of $r$ only, and Eq.~(\ref{32}) reduces to
\begin{align}\label{33}
    &\frac{e^{\psi-\lambda}}{r^4}\frac{d}{dr}\left[ e^{-(\psi+\lambda)}r^4\frac{d\varpi}{dr} \right] = 4(4\pi+ \beta)(\rho+ p_t)\varpi ,
\end{align}
and taking into account that $e^{-(\psi+\lambda)}= 1$ at the edge of the star and beyond, the last equation can be integrated to give
\begin{equation}\label{34}
    \left[ r^4\frac{d\varpi}{dr} \right]_{r_{\rm sur}} = 4(4\pi+ \beta)\int_0^{r_{\rm sur}}(\rho+ p_t)r^4e^{\lambda-\psi}\varpi dr . 
\end{equation}

From Eq.~(\ref{34}) we can obtain the relativistic moment of inertia of a slowly rotating anisotropic compact star in $f(R,T) = R+ 2\beta T$ gravity by means of expression 
\begin{equation}\label{35}
    I = \frac{2}{3}(4\pi+ \beta)\int_0^{r_{\rm sur}}(\rho+ p_r+ \sigma)e^{\lambda-\psi}r^4 \left( \frac{\varpi}{\Omega} \right)dr ,
\end{equation}
and hence the angular momentum $J= I\Omega$ can be written as
\begin{equation}\label{36}
 J = \frac{2}{3}(4\pi+ \beta)\int_0^{r_{\rm sur}} \frac{\rho+ p_r+ \sigma}{\sqrt{1- 2m/r}} (\Omega - \omega)e^{-\psi}r^4 dr .
\end{equation}

It can be seen that the above result then reduces to the pure general relativistic expression when $\beta= 0$. Furthermore, when both parameters $\beta$ and $\sigma$ vanish, Eq.~(\ref{36}) reduces to the expression given in Ref.~\cite{Glendenning} for isotropic compact stars in Einstein gravity. Analogously as in GR, the differential equation (\ref{33}) will be integrated from the origin at $r=0$ with an arbitrary choice of the central value $\varpi(0)$ and with vanishing slope, i.e., $d\varpi/dr=0$. Once the solution for $\varpi(r)$ is found, we can then compute the moment of inertia via the integral (\ref{35}).


\section{Equation of state and anisotropy ansatz}\label{Sec4}

Just as the construction of anisotropic compact stars in GR, to close the system of Eqs.~(\ref{22})-(\ref{24}), one needs to specify a barotropic EoS (which relates the radial pressure to the mass density by means of equation $p_r = p_r(\rho)$) and also assign an anisotropy function $\sigma$ since there is now an extra degree of freedom $p_t$. Alternatively, it is possible to assign an EoS for radial pressure and another for tangential pressure. For instance, an approach for the study of anisotropic fluids has been recently carried out within the context of Newtonian gravity in Ref.~\cite{Abellan2020a} and in conventional GR \cite{Abellan2020b}, where both the radial and tangential pressures satisfy a polytropic EoS.

In this work, we will follow the first procedure described in the previous paragraph in order to deal with anisotropic neutron stars within the framework of $f(R,T)$ gravity. Indeed, for radial pressure we use a well-known and physically relevant EoS which is compatible with the constraints of the GW170817 event (the first detection of gravitational waves from a binary neutron star inspiral \cite{Abbott}), namely, the soft SLy EoS \cite{DouchinHaensel2001}. This EoS is based on the SLy effective nucleon-nucleon interaction, which is suitable for the description of strong interactions in the nucleon component of dense neutron-star matter. Such unified EoS describes both the neutron-star crust and the liquid core (which is assumed to be a ``minimal'' $\rm npe\mu$ composition), and it can be represented by the following analytical expression
\begin{align}\label{SLyEoS}
  \zeta(\xi) =& \ \frac{a_1 + a_2\xi + a_3\xi^3}{1 + a_4\xi}f(a_5(\xi - a_6))  \nonumber  \\
  &+ (a_7 + a_8\xi)f(a_9(a_{10} - \xi))   \nonumber  \\
  &+ (a_{11} + a_{12}\xi)f(a_{13}(a_{14} - \xi))  \nonumber  \\
  &+ (a_{15} + a_{16}\xi)f(a_{17}(a_{18} - \xi)) ,
\end{align}
where $\zeta \equiv \log(p_r/ \rm dyn\ cm^{-2})$, $\xi \equiv \log(\rho/ \rm g\ cm^{-3})$, and $f(x) \equiv 1/(e^x +1)$. The values $a_i$ are fitting parameters and can be found in Ref.~\cite{HaenselPotekhin2004}.

In addition, we adopt the anisotropy ansatz proposed by Horvat \textit{et al.} \cite{Horvat2011} to model anisotropic matter inside compact stars, namely
\begin{equation}\label{ansatz}
    \sigma = \alpha p_r\mu = \alpha p_r(1- e^{-2\lambda}) ,
\end{equation}
with $\mu(r) \equiv 2m/r$ being the compactness of the star. The advantage of this ansatz is that the stellar fluid becomes isotropic at the origin since $\mu \sim r^2$ when $r \rightarrow 0$. It is also commonly known as quasi-local ansatz in the literature \cite{Horvat2011}, where $\alpha$ controls the amount of anisotropy inside the star and in principle can assume positive or negative values \cite{Pretel2020EPJC, Horvat2011, Folomeev2018, Pretel2022, Silva2015, Doneva2012, Yagi2015}. Note that in the Newtonian limit, when the pressure contribution to the energy density is negligible, the effect of anisotropy vanishes in the hydrostatic equilibrium equation. Regardless of the particular functional form of the anisotropy model, here we must emphasize that physically relevant solutions correspond to $p_r, p_t \geq 0$ for $r \leq r_{\rm sur}$.


\section{Numerical results and discussion}\label{Sec5}

\begin{figure*}
 \includegraphics[width=8.8cm]{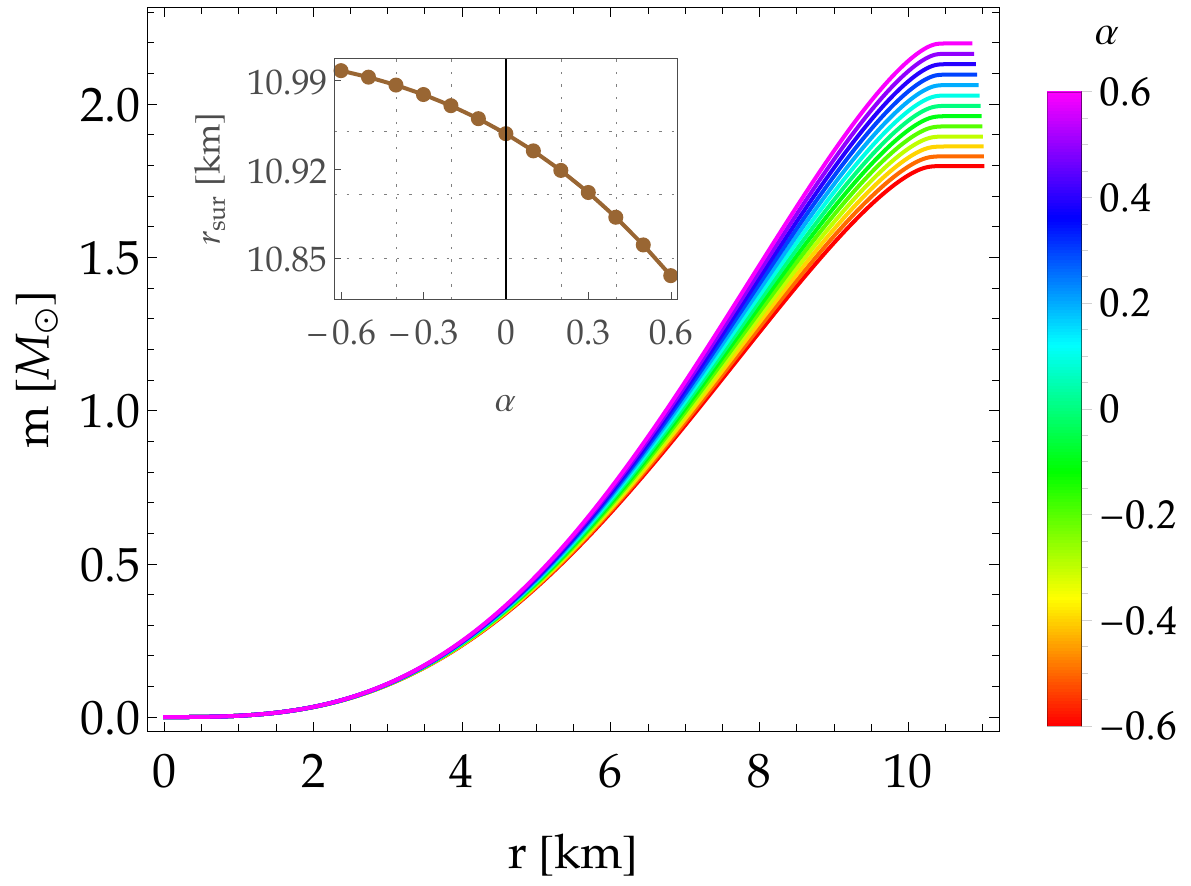}
 \includegraphics[width=8.8cm]{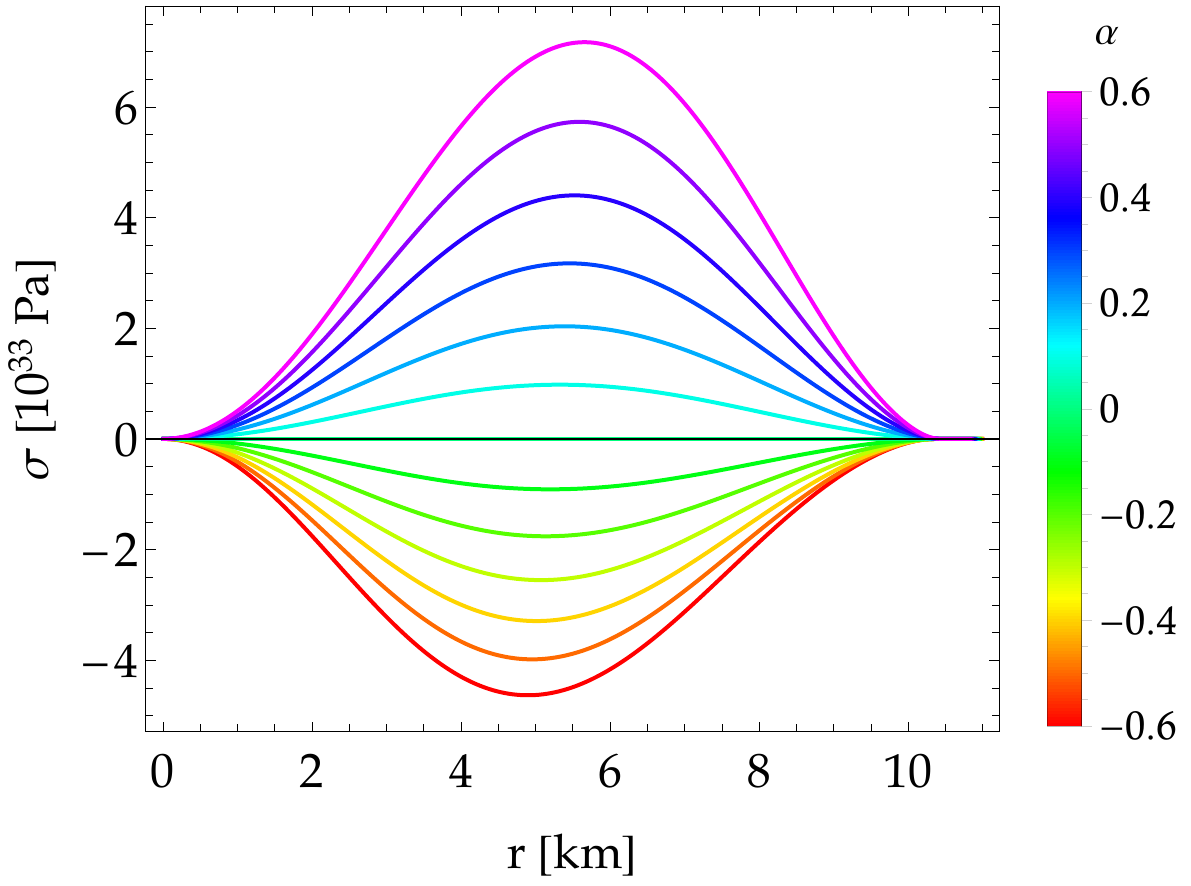}
 \caption{\label{figure1} Radial behaviour of the mass function (left panel) and the anisotropy factor (right panel) in the framework of $f(R,T)= R+2\beta T$ gravity for $\beta= -0.01$ and different values of $\alpha$. SLy EoS (\ref{SLyEoS}) is valid from $10^{11}\, \rm kg/m^3$ up to the maximum density reachable within neutron stars \cite{DouchinHaensel2001}, and in these plots we have considered $\rho_c = 2.0 \times 10^{18}\, \rm kg/m^3$. The isotropic case is recovered when the anisotropy parameter vanishes (this is, $\alpha = 0$). We can observe that the gravitational mass increases and the radius decreases as $\alpha$ increases. In addition, the anisotropy is more pronounced in the intermediate regions and vanishes at the stellar center as expected. }  
\end{figure*}

Given an EoS for the radial pressure, we numerically integrate the modified TOV equations (\ref{22})-(\ref{24}) with boundary conditions (\ref{25}) from the stellar center to the surface $r= r_{\rm sur}$ where the radial pressure vanishes. In addition, we have to specify a particular value for the coupling constant $\beta$ and for anisotropy parameter $\alpha$ which appears in Eq.~(\ref{ansatz}). For instance, for a central mass density $\rho_c = 2.0 \times 10^{18}\, \rm kg/m^3$ with SLy EoS (\ref{SLyEoS}), Fig.~\ref{figure1} illustrates the mass function and anisotropy factor as functions of the radial coordinate for $\beta= -0.01$ and several values of $\alpha$. The left plot reveals an increase in gravitational mass and a decrease in radius as $\alpha$ increases. Moreover, from the right plot we can see that the anisotropy vanishes at the center (which is a required condition in order to guarantee regularity), is more pronounced in the intermediate regions, and it vanishes again at the stellar surface.

\begin{figure*}
 \includegraphics[width=8.8cm]{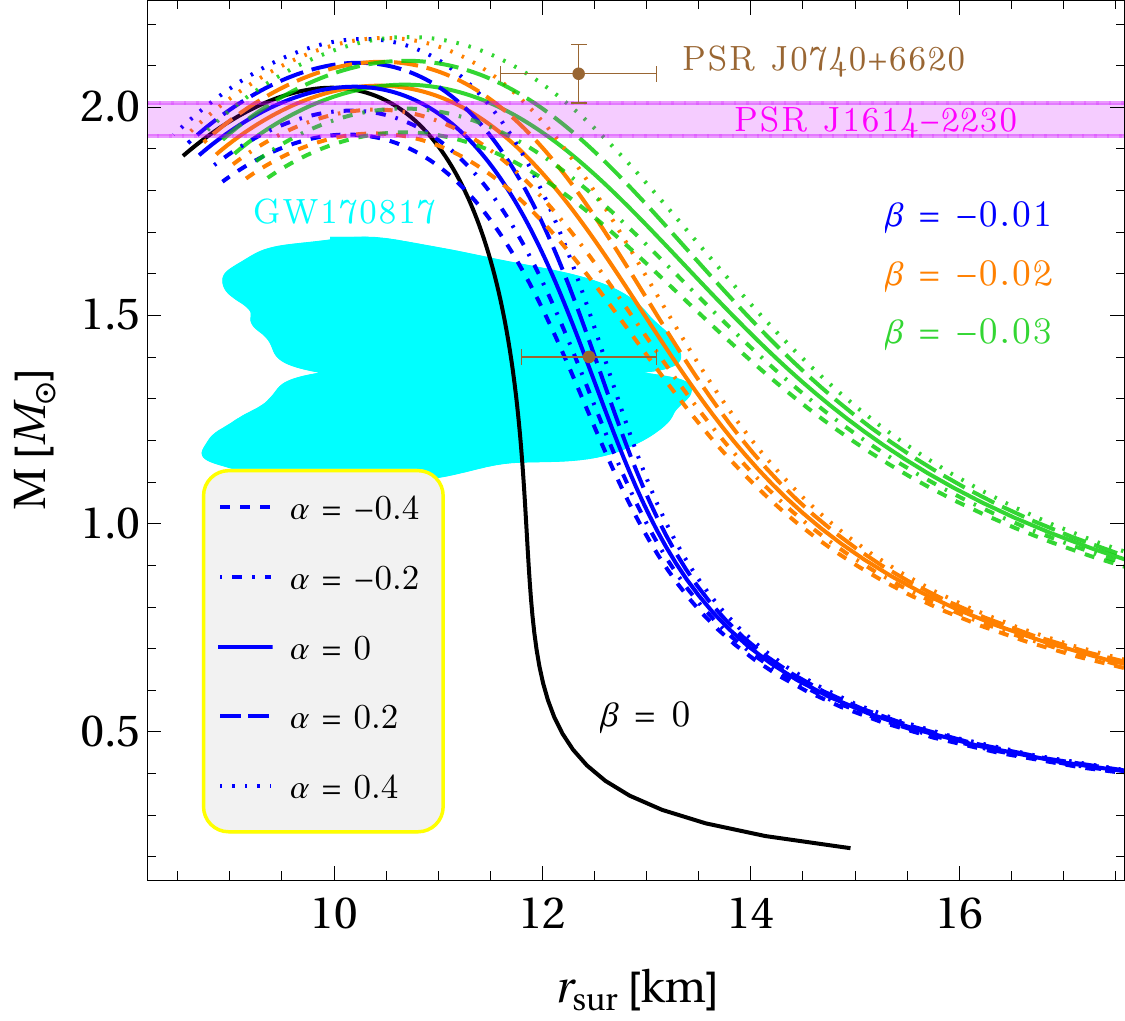} \
 \includegraphics[width=8.8cm]{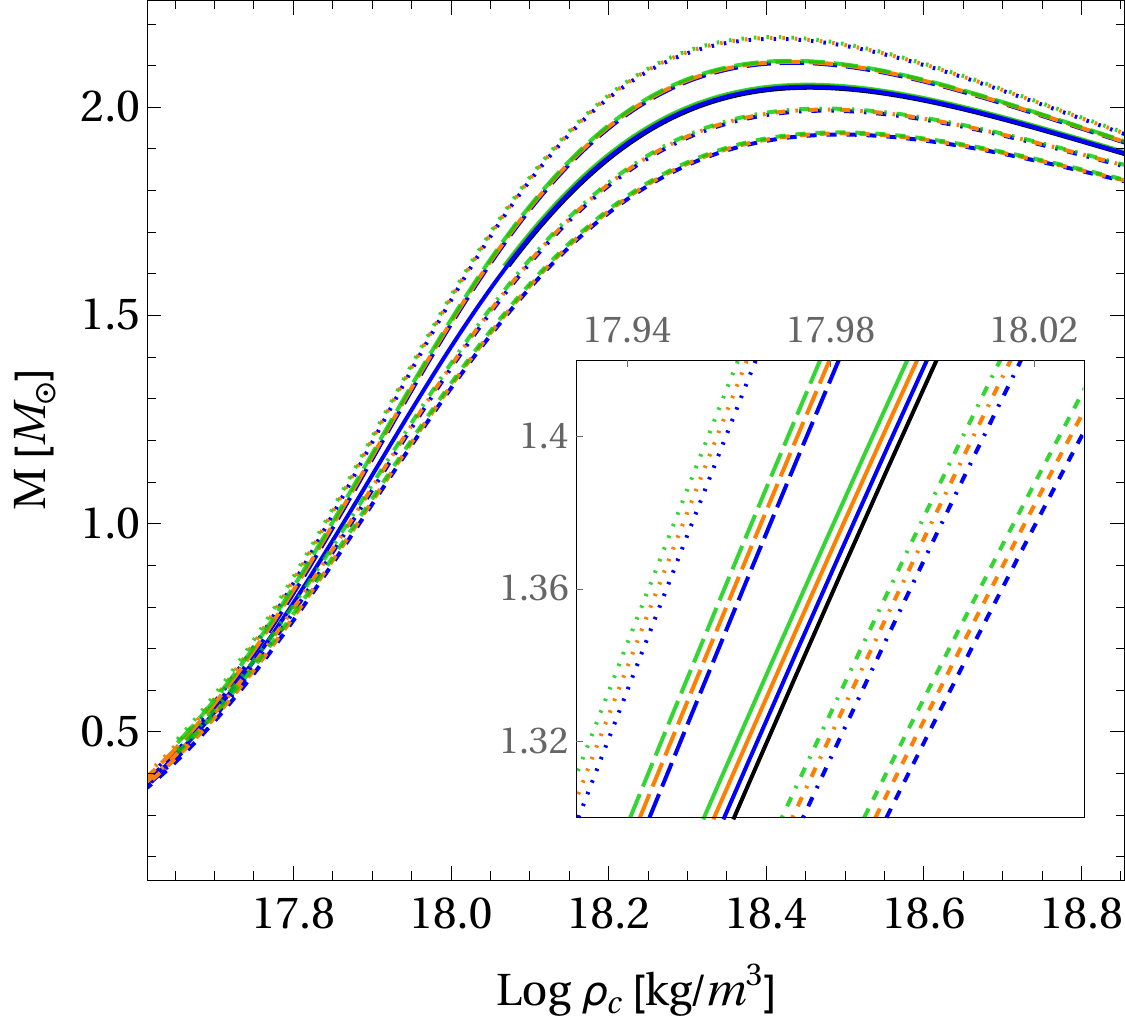}
 \caption{\label{figure2} Mass-radius diagrams (left panel) and mass-central density relations (right panel) for anisotropic neutron stars with SLy EoS (\ref{SLyEoS}) in $f(R,T)= R+2\beta T$ gravity for $\beta = -0.01$ (blue curves), $\beta= -0.02$ (orange curves) and $\beta = -0.03$ (in green). The solid lines correspond to $\alpha= 0$ (that is, isotropic solutions), and the pure GR case ($\beta= 0$) is shown in both plots as a benchmark by a black line. The magenta horizontal band stands for the observational measurement for the millisecond pulsar J1614-2230 reported in Ref.~\cite{Demorest2010}. The filled cyan region is the mass-radius constraint from the GW170817 event. The Radius of PSR J0740+6620 from NICER and XMM-Newton Data \cite{Miller2021} is indicated by the top brown dot with their respective error bars. Moreover, the bottom brown dot represents the radius estimate for a $1.4\, M_\odot$ neutron star \cite{Miller2021}. }  
\end{figure*}

For the anisotropy function (\ref{ansatz}), the left panel of Fig.~\ref{figure2} displays the mass-radius relations for anisotropic neutron stars with SLy EoS in $f(R,T)= R+2\beta T$ gravity for three particular values of the coupling constant $\beta$ and different values of $\alpha$. Here the total gravitational mass of each configuration is given by $M= m(r_{\rm sur})$, and the isotropic case in Einstein gravity has been included for comparison purposes by a black solid line. The mass-radius relation exhibits substantial deviations from GR mainly in the low-mass region. On the other hand, anisotropy introduces considerable changes only in the high-mass region. We remark that the $2\beta T$ term together with the presence of anisotropies (with positive values of $\alpha$) allow us to obtain maximum masses bigger than $2.0\, M_\odot$. As a consequence, the introduction of anisotropies in $f(R,T)= R+2\beta T$ gravity gives rise to massive neutron stars that are in good agreement with the millisecond pulsar observations \cite{Demorest2010, Cromartie2019}. From NICER and XMM-Newton data \cite{Miller2021}, the radius measurement for a $1.4\, M_{\odot}$ neutron star is $12.45 \pm 0.65\, \rm km$ and, according to the mass-radius diagram, our results consistently describe this star when $\beta = -0.01$ (see blue curves). Furthermore, it should be noted that the parameter $\beta=-0.01$ is the one that best fits the mass-radius constraint from the GW170817 event (see the filled cyan region). Nevertheless, the massive pulsar J0740+6620 (whose radius is $12.35 \pm 0.75\, \rm km$ \cite{Miller2021}) could be described only when $\beta= -0.03$ and $\alpha = 0.4$.

It is worth commenting that the value of the parameter $\alpha$ could be constrained, but that will depend on the particular compact star observed in the Universe. For instance, the range $\alpha \in [-0.4, -0.2]$ consistently describes the millisecond pulsar J1614-2230 regardless of the value of $\beta$. However, for highly massive neutron stars whose masses are greater than $2.0\, M_\odot$, positive values of $\alpha$ will be required. For PSR J0740+6620, whose gravitational mass is $2.08\, M_{\odot}$, the best value for $\alpha$ is 0.2. In fact, this constraint will depend not only on the modified theory of gravity but also on the equation of state adopted for the radial pressure.

\begin{figure*}
 \includegraphics[width=8.8cm]{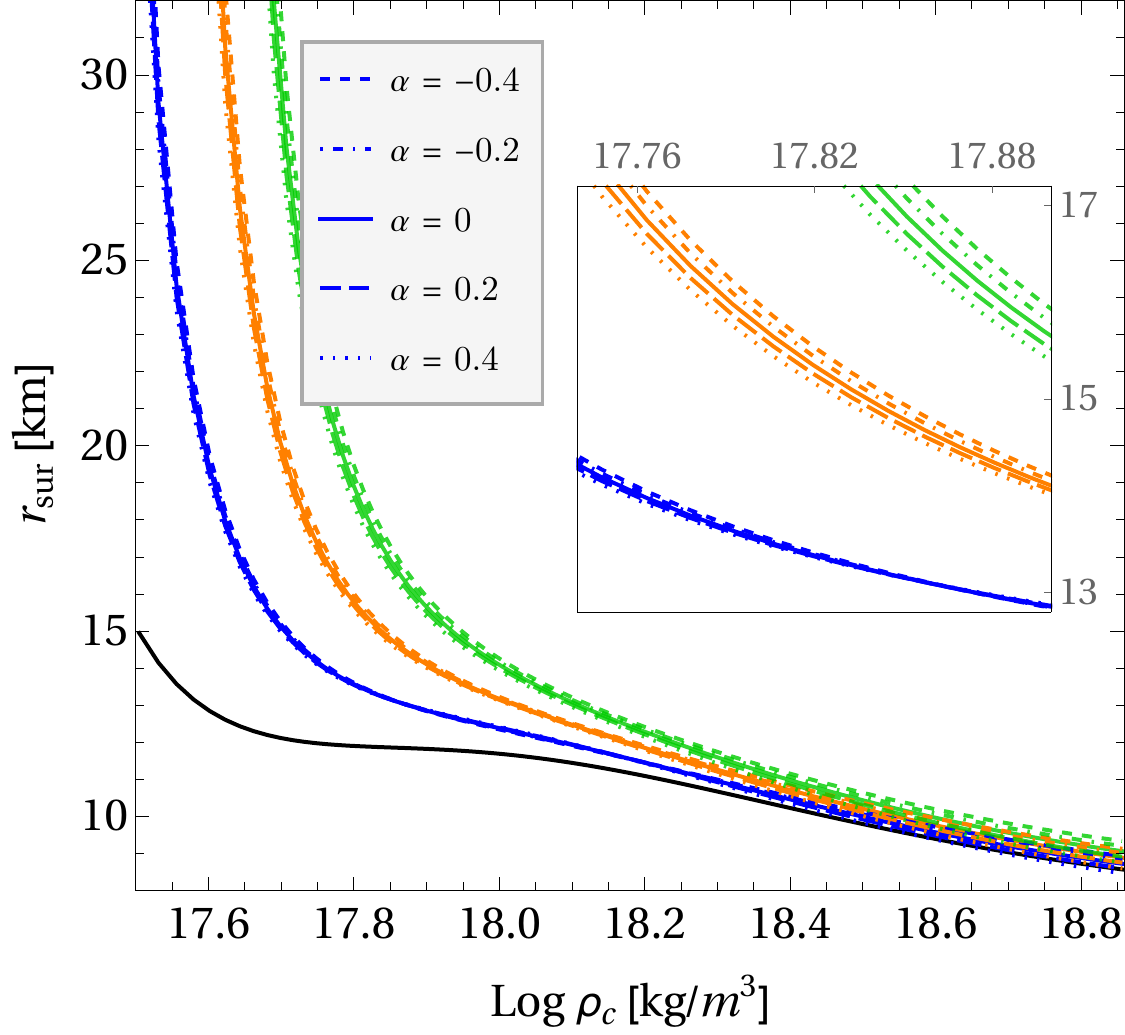} \
 \includegraphics[width=8.8cm]{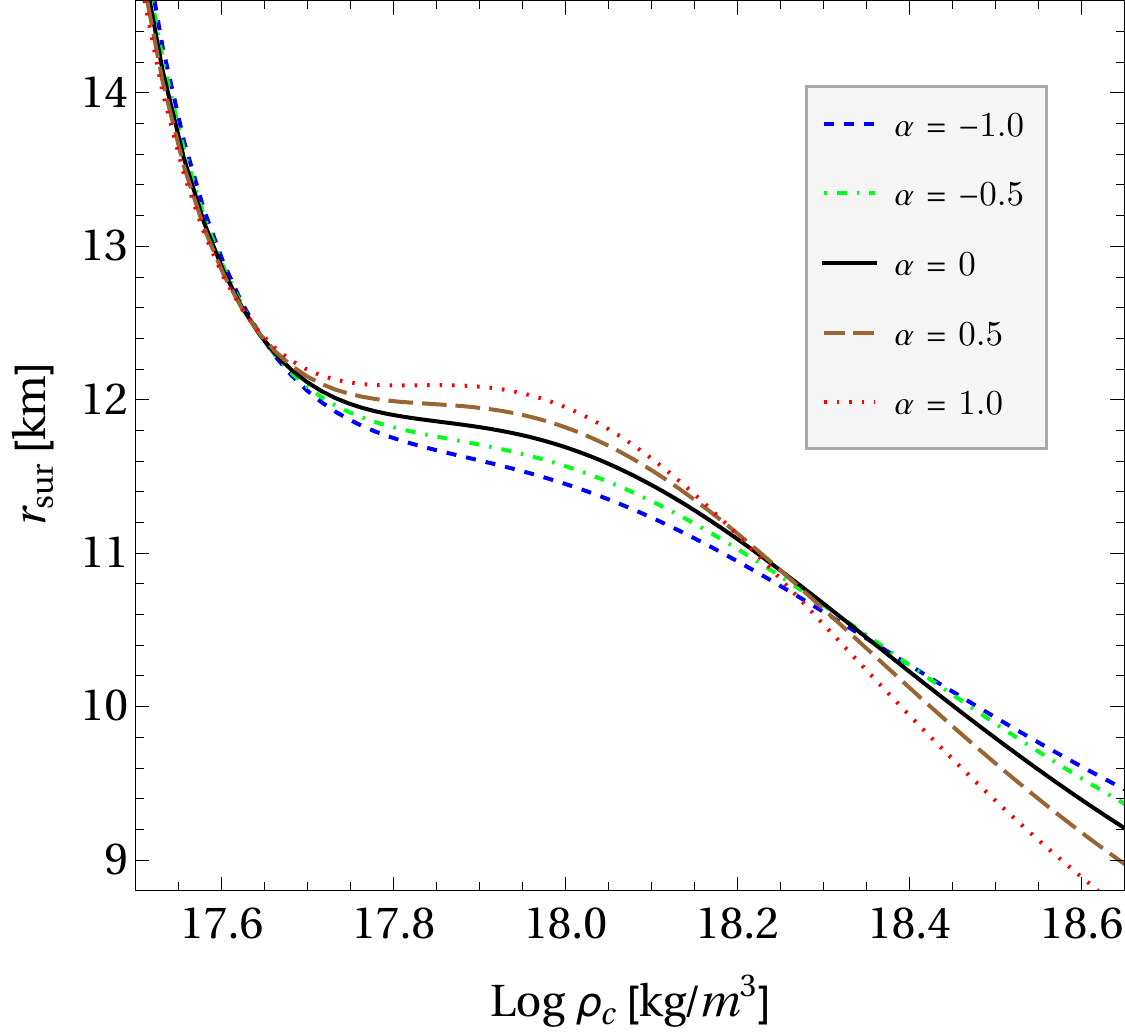}
 \caption{\label{figure3} Surface radius as a function of the central mass density. On the left panel, different styles and colors of the curves correspond to different values of the parameters $\beta$ and $\alpha$ as in Fig.~\ref{figure2}. The most substantial deviations from GR take place at low central densities, whereas for large central densities the changes are very slight due to the $2\beta T$ term. On the right panel we display the modifications of the radius due to the inclusion of anisotropies when $\beta= 0$, where we have considered larger values for $\vert\alpha\vert$ in order to appreciate the changes in radius as a consequence of anisotropy. We can mainly observe three regions where the radius can decrease or increase depending on the value of $\alpha$. }
\end{figure*}

According to the right panel of Fig.~\ref{figure2}, the parameter $\beta$ slightly modifies the total gravitational mass, however, the effect of anisotropy introduces more relevant changes. To better analyze the effects that arise as a result of the modification of Einstein's theory as well as the incorporation of anisotropies, in Fig.~\ref{figure3} we show the behavior of the surface radius as a function of the central density. From the left plot we can conclude that the radius is significantly altered due to the $2\beta T$ term in the low-central-density region, while anisotropy slightly modifies the radius of the stars. The right plot corresponds to the pure general relativistic case and it can be observed that the radius undergoes more significant modifications with respect to its isotropic counterpart if the values for $\vert\alpha\vert$ are larger than those considered in the left plot.

\begin{figure*}
 \includegraphics[width=8.8cm]{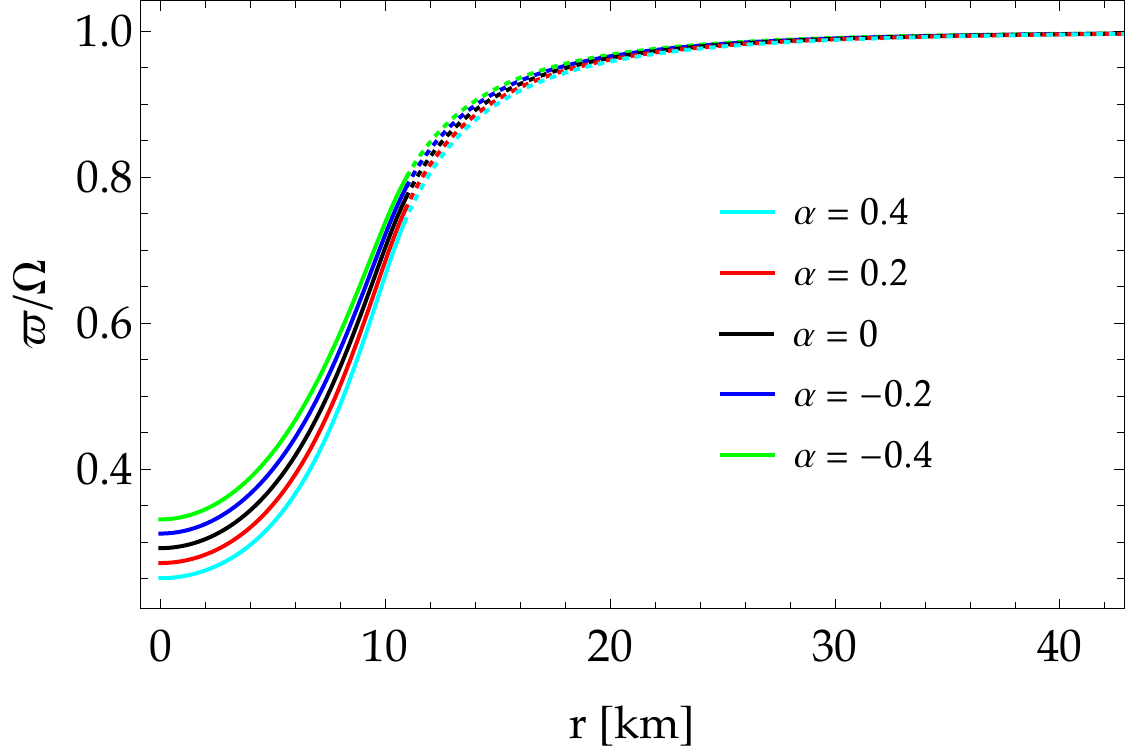} \
 \includegraphics[width=8.8cm]{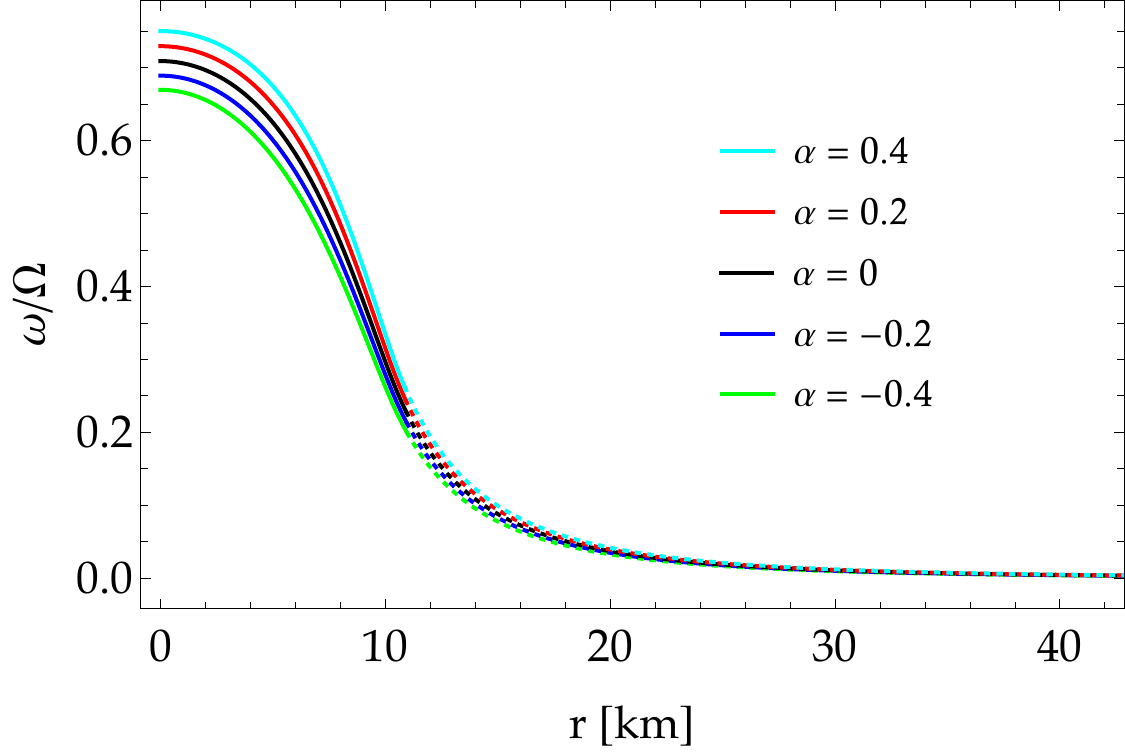}
 \caption{\label{figure4} Left panel: Numerical solution of the differential equation (\ref{33}) for a given central mass density $\rho_c = 2.0 \times 10^{18}\, \rm kg/m^3$ in $f(R,T)= R+2\beta T$ gravity with $\beta= -0.01$ and different values of the free parameter $\alpha$. The dotted lines represent the solutions of the exterior region, and as expected $\varpi \rightarrow \Omega$ at great distances from the stellar surface. Right panel: Ratio of frame-dragging angular velocity to the angular velocity of the stars, namely $\omega(r)/\Omega = 1- \varpi(r)/\Omega$. Notice that the solution of the exterior problem provides an asymptotic behavior of $\omega(r)$. }
\end{figure*}

Eq.~(\ref{33}) is first solved in the interior region from the center to the surface of the star by considering an arbitrary value for $\varpi$ and with vanishing slope at $r=0$. Then the same equation is solved in exterior spacetime from the surface to a sufficiently far distance from the star where $\varpi(r) \rightarrow \Omega$. In Fig.~\ref{figure4} we display the radial profile of these solutions for the central mass density considered above. We observe that $\varpi(r)$ is an increasing function of the radial coordinate, whereas $\omega(r)$ is a decreasing function and hence the largest rate of dragging of local inertial frames always occurs at the stellar center. Furthermore, appreciable effects (mainly in the interior region of the stellar configuration) can be noted on frame-dragging angular velocity due to the inclusion of anisotropies. 

\begin{figure*}
 \includegraphics[width=8.8cm]{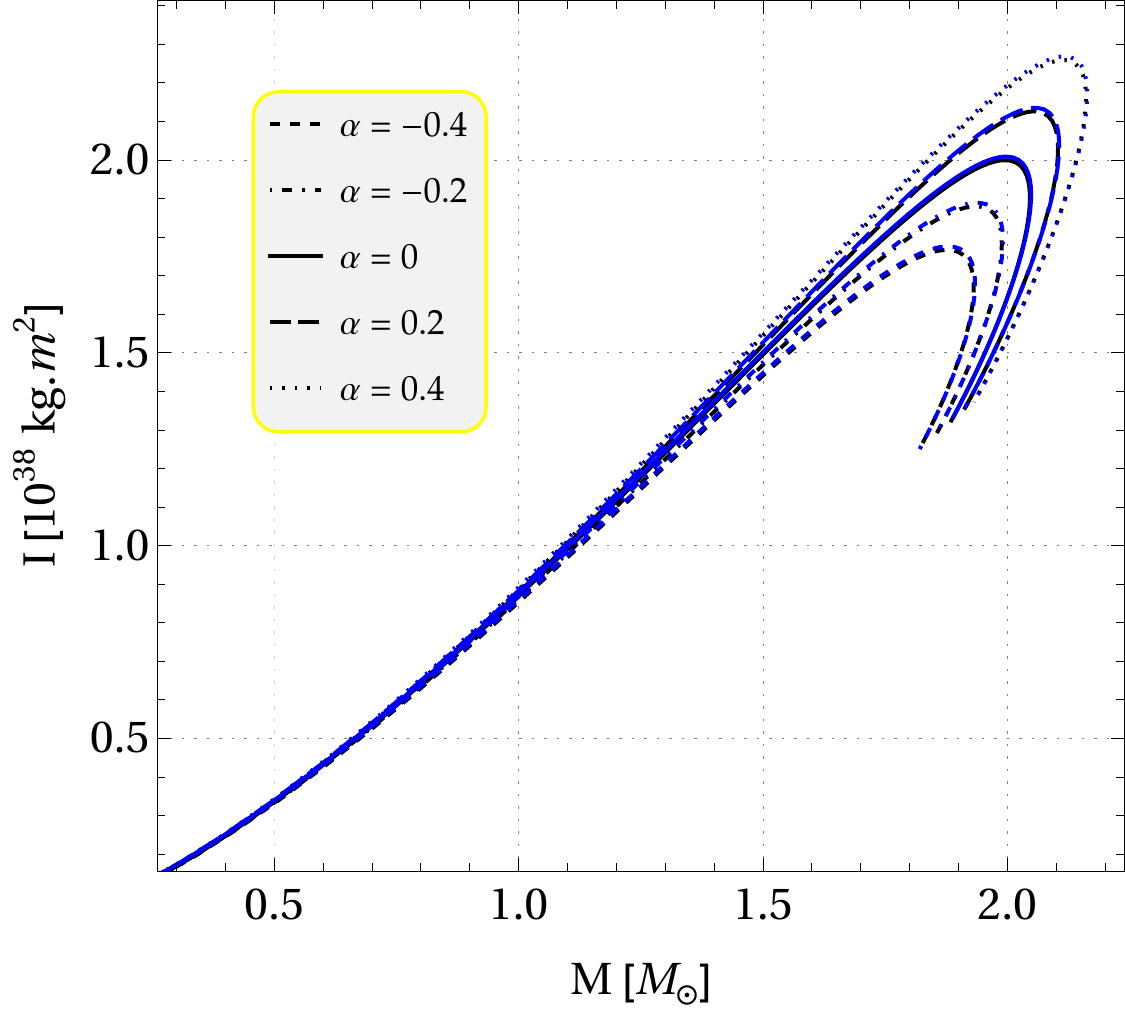} \
 \includegraphics[width=8.8cm]{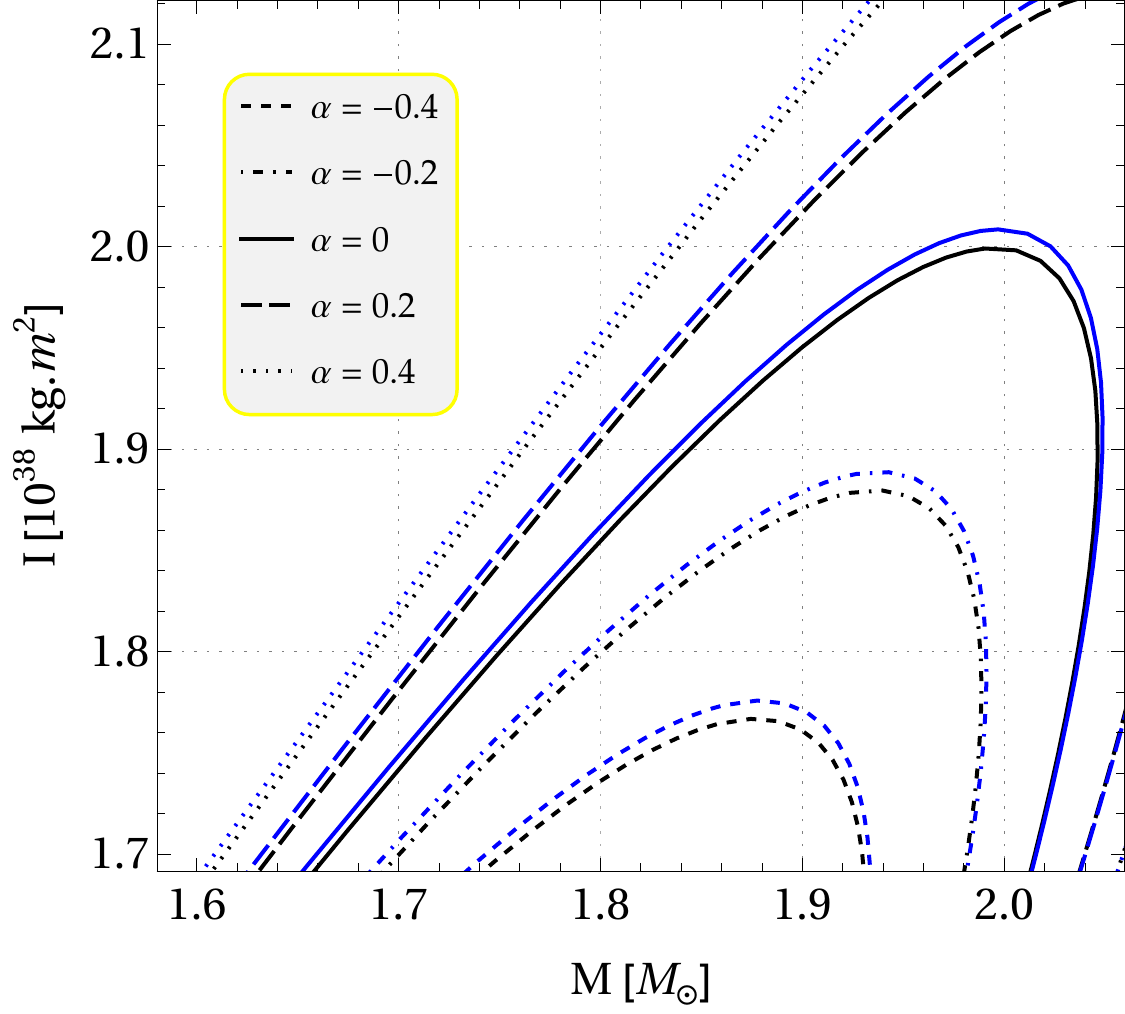}
 \caption{\label{figure5} Left panel: Moment of inertia of slowly rotating anisotropic neutron stars as a function of the total mass within the context of $f(R,T)= R+2\beta T$ gravity for $\beta = -0.01$ in blue. Different styles of the curves correspond to different values of the anisotropy parameter $\alpha$. Results based on Einstein's theory have been included for comparison purposes and are represented by the black curves. We can appreciate that the moment of inertia is modified very slightly by the $2\beta T$ term, however, the anisotropies introduce relevant changes in the large-mass region. The right plot is a magnification of the left one. }
\end{figure*}

Once $\varpi(r)$ is known for each stellar configuration, we can then determine the moment of inertia by means of Eq.~(\ref{35}). Figure \ref{figure5} presents the moment of inertia as a function of the total gravitational mass in GR and within the context of $f(R,T)= R+2\beta T$ gravity for $\beta= -0.01$. It can be observed that the moment of inertia undergoes irrelevant changes from GR, however, it can change significantly due to anisotropies in the high-mass region.

\section{Conclusions}\label{Sec6}

In this work we have investigated slowly rotating anisotropic neutron stars in $f(R,T)= R+2\beta T$ gravity, where the degree of modification with respect to GR is measured by the coupling constant $\beta$. The modified TOV equations and moment of inertia have been derived within the context of anisotropic fluids by retaining only first-order terms in the angular velocity as measured by a distant observer ($\Omega$). Notice that, within this linear approximation, the moment of inertia can be calculated from the structure of a non-rotating configuration since the TOV equations describing the static background are still valid. In addition, we have adopted the anisotropy ansatz proposed by Horvat and collaborators \cite{Horvat2011}, where appears a dimensionless parameter $\alpha$ which measures the degree of anisotropy within the neutron star.

We have analyzed the consequences of the extra term $2\beta T$ together with anisotropies on the properties of neutron stars such as radius, mass, frame-dragging angular velocity and moment of inertia. Indeed, our results reveal that the radius deviates considerably from GR in the low-central-density region, however, the total gravitational mass and the moment of inertia undergo slight modifications due to the influence of the effects generated by the minimal matter-gravity coupling. Furthermore, the presence of anisotropy generates substantial changes both in the mass and in the moment of inertia with respect to the isotropic case. The appreciable effects due to the inclusion of anisotropy occur mainly in the higher-central-density region, this is, for large masses (near the maximum-mass configuration).

\begin{acknowledgments}
JMZP acknowledges financial support from the PCI program of the Brazilian agency ``Conselho Nacional de Desenvolvimento Científico e Tecnológico''--CNPq. 

\end{acknowledgments}\


%

\end{document}